\documentclass[runningheads,a4paper]{llncs}

\usepackage{amssymb,amsmath}

\usepackage{arydshln}

\usepackage{graphicx}

\usepackage[breaklinks,colorlinks=true,citecolor=blue]{hyperref}

\DeclareMathOperator*{\argmax}{arg\,max}

\begin{document}

\mainmatter  % start of an individual contribution

\title{``Infographics'' team: Selecting Control Parameters via Maximal Fisher Information}

%\titlerunning{Gliders2014}

\author{Siddharth Pritam}

\institute{Alumni Group of Indian Institute of Technology\\
Kharagpur, India\\
email: sid.iitkgp@gmail.com
}

%\toctitle{Lecture Notes in Computer Science}
%\tocauthor{Authors' Instructions}

\maketitle
\setcounter{footnote}{0}

%
%\begin{abstract}
%
%\end{abstract}
%

\section{Introduction}

 ``Infographics''  is a very new RoboCup Soccer Simulation League 2D team, based on the released source code of agent2d \cite{agent2d} and  Gliders \cite{POWH12,gliders2013tdp}. It also uses librcsc (a base library for The RoboCup Soccer Simulator, RCSS \cite{Manual2002}); soccerwindow2: a viewer program for RCSS; fedit2: a formation editor for agent2d; and GIBBS: in-browser log-player for 2D Simulation League logs \cite{moore}.

The Soccer Simulation 2D League is one of the oldest RoboCup leagues which still presents a significant challenge for Artificial Intelligence, Machine Learning, Autonomous Agents and Multi-Agent Systems, Cognitive Robotics and Complex Systems \cite{Noda2003,balch,IAAI2000,Butler2001,Reis2001,Prokopenko02,Prokopenko03}.  

In this  paper we describe an approach to utilize Fisher information in interaction networks, produced by dynamics of RoboCup Simulation games. This approach is based on (a) methods for computing interaction networks, introduced in \cite{cliff2013towards} and verified  by Gliders \cite{gliders2013tdp}, and (b) methods for determining Fisher information in complex networks, described in \cite{PLOW11}.  It guides optimization of control parameters used by ``Infographics''.

\section{Background}

Information theory \cite{mac03} is an increasingly popular framework for the study of complex systems and their phase transitions \cite{ForemanPW03,liz08c,pro09,Prokopenko13}. In part, this is because complex systems can be viewed as distributed computing systems, and information theory is a natural way to study computation, e.g. \cite{liz08a,liz14a}. Information theory is applicable to any system, provided that one can define probability distribution functions for its states. 

One particular information-theoretic measure, Fisher information, has been extensively used in many fields of science, providing insights that can be directly compared across different system types. It measures the amount of information that an observable random variable has about an unknown parameter, and can also be intuitively interpreted as a measure of the sensitivity of the random variable to changes in the parameter.  For example, a recent study \cite{PLOW11} explicitly related elements of the Fisher information matrix to the rate of change in the
corresponding order parameters, and identified, in particular, second-order phase transitions via divergences of individual elements of the Fisher information matrix.

In parallel, our recent research has also focused on phase transitions between ordered and chaotic behaviour, analyzed from the perspective of distributed computation: specifically, in
small-world Random Boolean Networks (RBN's) \cite{liz11b,liz11b2}. These studies characterized the distributed computation in terms of its
underlying information dynamics: information storage and information transfer, and gave us a good insight on the behavior of the networks and flow of information near critical regimes.

Furthermore, in the context of RoboCup and team sports in general, several new techniques were recently developed for quantifying dynamic interactions in simulated football games \cite{cliff2013towards}. This study not only involved computation of information transfer and storage, but also related the information transfer to responsiveness of the players,
and the information storage within the team to the team's rigidity and lack of tactical flexibility \cite{cliff2013towards}.

We intend to bring these results together, by computing Fisher information within the interaction
networks obtained by the techniques introduced in \cite{cliff2013towards}, and characterising critical behaviour and
phase transitions via Fisher information, following methods of \cite{PLOW11}.
The results promise to advance research in complex multi-agent systems, as well as contribute to
development of new behaviours, implemented within our RoboCup team, ``Infographics'' (India).

\section{Methods}

The first step (1) of the proposed algorithm is computation of ``information-base diagram'', i.e. interaction networks \cite{cliff2013towards}.  It has the following sub-steps, reproduced from \cite{cliff2013towards}:

(1.1) The average conditional transfer entropy \cite{schr00} between any two agents from opposing teams $X$ and $Y$ is computed for a game $g$ with $N$ time steps:
\begin{equation}
    T^g_{Y_i \rightarrow X_j | B}  =  \frac{1}{N} \sum_{n=0}^{N-1} \lim_{k \to \infty} \log_{2}{ \frac{ p(x_{n+1} \mid x^{(k)}_{n},y_{n}, b_n)}{p(x_{n+1} \mid x^{(k)}_{n}, b_n)}} \ ,
\end{equation}
where $Y_i$ is the process tracing the change in the position $y_n$ of agent $i$ ($2 \leq i \leq 11$) from team $Y$ at time $n$ ($1 \leq n \leq N$); $X_j$ is the process tracing the change $x_n$ in the position of  agent $j$ ($2 \leq j \leq 11$) from team $X$ at time $n$; $B$ is the process tracing the change in current ball position $b_n$; and $x^{(k)}_{n} = \{ x_{n-k+1}, \ldots, x_{n-1}, x_n\}$ is the  past of the process $X_j$ (semi-infinite past as $k \to \infty $).

(1.2) The responder agent $J_X(Y_i)$ in team $X$ is identified as the mode of the series 
\[ \{ J_X(Y_i,1), \ldots, J_X(Y_i,G) \} \ , \]
for $G$ games, where
\begin{equation}
  J_X(Y_i,g) =  \argmax_{2 \ \leq \ j \ \leq \ 11}{ \ T^g_{Y_i \rightarrow X_j | B}} \ .
\end{equation}

(1.3) The pairs $(i, J_X)$, where $2 \leq i \leq 11$, identified for each agent $i$ in team $Y$, produce an ``information-base diagram'' $D(Y,X)$ \cite{cliff2013towards}.  There is one outgoing link from each agent $i$ in team $Y$, but there may be several incoming links to the agent $i$ from team $Y$, or an agent $j$ from team $X$.  At this sub-step we select the agent $\hat{J}_X$ in our team $X$ which has the maximal number of incoming links.  In other words, it is the responder agent to the maximal number of opponent agents, being a ``hub'' of the ``information-base diagram''.

\

The next step (2) is computation of Fisher information $F(X_{\hat{J}}|\theta)$ of time series $X_{\hat{J}}$, i.e., the process tracing the change in the hub-agent's position, with respect to some control parameter, $\theta$.  The control parameter may be chosen from a broad set of parameters used within the team.   Varying the control parameter within its range $\Theta$ allows us to compute Fisher information $F(X_{\hat{J}}|\theta)$ for different values of $\theta \in \Theta$, and select the value maximizing Fisher information:
\begin{equation}
\theta^* = \argmax_{\theta \in \Theta} F(X_{\hat{J}}|\theta) \ .
\end{equation}
As argued in \cite{PLOW11}, the value maximizing Fisher information corresponds to a critical point, pinpointing a phase transition.

In summary, this method leads to selection of critical values of control parameters, assisting in optimizing influence of hub-agents on the oppponent team.

\section{Conclusion}

The main motivation behind our approach is the study of phase transitions and relevant control parameters, using interaction networks of RoboCup teams and Fisher information.

We described a method for a selection of control parameters, driven by maximizing Fisher information of a hub-agent in an information-base diagram.  The mechanism is utilized by ``Infographics'', our new simulated soccer team which intends to participate in the RoboCup  Simulation 2D League in 2014.

\subsubsection*{Acknowledgments.}

We thank Hidehisa Akiyama and his colleagues for development of agent2d;  Mikhail Prokopenko, Peter Wang,  and Oliver Obst for their advices regarding the source code of Gliders and helpful discussions of this manuscript; as well as Joseph Lizier for assistance with his Java Information Dynamics toolkit.

\bibliographystyle{splncs}

%\bibliography{refs,references,TE}

\end{document}